\documentstyle[12pt]{article}

\input{psfig}

\font\mybb=msbm10 at 11pt
\def\bb#1{\hbox{\mybb#1}}
\def\e{{\rm e}}
\def\slash{\!\!\!\!/}
\def\tr{{\rm tr}}
\def\beq{\begin{equation}}
\def\eeq{\end{equation}}
\def\bea{\begin{eqnarray}}
\def\eea{\end{eqnarray}}
\def\bd{\begin{displaymath}}
\def\ed{\end{displaymath}}

\makeatletter
\newdimen\normalarrayskip              
\newdimen\minarrayskip                 
\normalarrayskip\baselineskip
\minarrayskip\jot
\newif\ifold             \oldtrue            \def\new{\oldfalse}
\def\arraymode{\ifold\relax\else\displaystyle\fi} 
\def\@arrayskip{\ifold\baselineskip\z@\lineskip\z@
     \else
     \baselineskip\minarrayskip\lineskip2\minarrayskip\fi}
\def\@arrayclassz{\ifcase \@lastchclass \@acolampacol \or
\@ampacol \or \or \or \@addamp \or
   \@acolampacol \or \@firstampfalse \@acol \fi
\edef\@preamble{\@preamble
  \ifcase \@chnum
     \hfil$\relax\arraymode\@sharp$\hfil
     \or $\relax\arraymode\@sharp$\hfil
     \or \hfil$\relax\arraymode\@sharp$\fi}}
\def\@array[#1]#2{\setbox\@arstrutbox=\hbox{\vrule
     height\arraystretch \ht\strutbox
     depth\arraystretch \dp\strutbox
     width\z@}\@mkpream{#2}\edef\@preamble{\halign \noexpand\@halignto
\bgroup \tabskip\z@ \@arstrut \@preamble \tabskip\z@ \cr}%
\let\@startpbox\@@startpbox \let\@endpbox\@@endpbox
  \if #1t\vtop \else \if#1b\vbox \else \vcenter \fi\fi
  \bgroup \let\par\relax
  \let\@sharp##\let\protect\relax
  \@arrayskip\@preamble}
\makeatother

\newlength{\extraspace}
\newlength{\extraspaces}
\begin{document}

\renewcommand{\footnotesize}{\small}

\addtolength{\baselineskip}{.8mm}

\thispagestyle{empty}

\begin{flushright}
{\sc OUTP}-97-15P\\
  hep-th/9703209\\
\hfill{  }\\ March 28, 1997
\end{flushright}
\vspace{.3cm}

\begin{center}
{\Large\sc{\bf Matrix Strings \\in Two-dimensional Yang-Mills Theory}}\\[15mm]

{\sc  Ian I. Kogan and Richard J.\ Szabo}\footnote{Work supported in part by
the Natural Sciences and Engineering Research Council of Canada.} \\[2mm]
{\it Department of Physics -- Theoretical Physics\\ University of Oxford\\ 1
Keble Road, Oxford OX1 3NP, U.K.} \\[15mm]

\vskip 1.5 in

{\sc Abstract}

\begin{center}
\begin{minipage}{14cm}

We describe the structure of string vacuum states in the supersymmetric matrix
model for M theory compactified on a circle in the large-$N$ limit. We show
that the theory admits topological instanton field configurations which at
short-distance scales reduce to ordinary Yang-Mills instantons that interpolate
between degenerate vacua of the theory. We show that there exists further
classes of hadronic strings associated with the D-string super-fields. We
discuss the relationships between these non-perturbative string states and
rigid QCD strings, critical strings, and membrane states.

\end{minipage}
\end{center}

\end{center}

\noindent

\vfill
\newpage
\pagestyle{plain}
\setcounter{page}{1}

`M theory' is the non-perturbative theory that is a quantum extension of
11-dimensional supergravity \cite{wittenM}. Via compactification on a circle
$S^1$, it is equivalent to 10-dimensional type-IIA superstring theory. M theory
in this way subsumes all five consistent superstring theories in 10 dimensions
and contains 11-dimensional supergravity
as its low-energy limit. It has been suggested recently that the full
non-perturbative dynamics of this theory lies in the content of an appropriate
large-$N$ limit of a supersymmetric matrix model \cite{bfss}. This
$N\times N$ matrix quantum mechanics is obtained as the dimensional reduction
of supersymmetric $U(N)$ Yang-Mills gauge theory in 10 dimensions
\cite{halpern} and it describes the low-energy collective dynamics of $N$
parallel D0-branes in the weak-coupling limit of type-IIA superstring theory
\cite{wittenD,dfs}. As each block $N_i\times N_i$ sub-matrix, with
$\sum_iN_i=N$, describes a non-relativistic super-membrane (regarded as a
collective excitation of D0-branes)  \cite{dhntown}, the $N\to\infty$ limit of
the matrix model describes a multi-membrane theory and thus naturally
incorporates the quantum field theoretical Fock space for the membranes of
supergravity. The permutation symmetry group $S_N$ of block diagonal matrices
is naturally realized as the Weyl group of the gauge group $U(N)$, and the
non-commutativity of the matrix coordinate fields at small distance scales
gives a realization of the idea that at distances smaller than the Planck
length the conventional notion of a spacetime geometry breaks down. Since the
longitudinal momentum in the 11-th direction is $p_{11}=N/R_{11}$, with
$R_{11}$ the compactification radius, the large-$N$ limit also effectively
accomplishes a decompactification of the extra 11-th direction and should
describe the short-distance space-time structure of M theory in the infinite
momentum frame. This conjecture has thus far survived a number of consistency
checks \cite{berdoug} leading to a ``third quantization" of string theory.

In this Letter we will study some features of the spectrum of states of the
supersymmetric matrix model for M theory when one of the nine transverse
directions is compactified on a circle $S^1$. In this case, the matrix quantum
mechanics becomes a two-dimensional matrix field theory coupled to
two-dimensional Yang-Mills theory which is also obtained as the dimensional
reduction of 10-dimensional supersymmetric Yang-Mills theory down to
two-dimensions, and it describes the non-relativistic dynamics of D-strings
\cite{bfss},\cite{taylor}--\cite{fundstring}. The non-perturbative formulation
of string theory as a two-dimensional supersymmetric Yang-Mills theory was
first proposed in \cite{motl}. This viewpoint was elaborated on in \cite{dvv}
where it was discussed how to identify the particle spectrum
of string theory with the states that can be made up from infinitely many
D-particles. In the following we shall study the particle spectrum of the
D-string field theory by considering the effective two-dimensional Yang-Mills
theory that is obtained. We shall see that the two-dimensional gauge theory
that we study in this way is an important model for deducing the spectrum
of string vacuum states, and also for a non-perturbative quantum gauge theory
framework for string theory. Using the $SO(8)$ triality property of the theory
\cite{kss} we show that in the large-$N$ limit the gauged action for the
D-string super-fields can be diagonalized by an explicit Nicolai mapping
\cite{nicolai,bbrt}. From this we identify some of the non-perturbative degrees
of freedom in the large-$N$ limit that describe the string vacuum states as
topological instanton fields from the string world-sheet into the
eight-dimensional target space. These instantons interpolate between ordinary
Yang-Mills instantons on the world-sheet at short-distance scales, which
connect the usual $\theta$-vacua of two-dimensional adjoint quantum
chromodynamics (QCD), and the D-string configurations at large-distance scales
which describe the free string limit of the theory. We emphasize the potential
relevance of the instanton sum that we find to the construction of
non-perturbative membrane states. We also demonstrate the appearence of certain
hadronic strings constructed from the D-string super-fields of the theory, and
we discuss the properties of these string states in conjuction with rigid QCD
strings and critical strings. We also discuss the potential interplay between
the instanton moduli space that we find and the conventional moduli spaces that
arise in two-dimensional Yang-Mills theory \cite{bbrt}.

The D-string field theory is the two-dimensional ${\cal N}=8$ supersymmetric
$U(N)$ Yang-Mills theory with action
\beq\new{\begin{array}{lll}
S[A;X,\psi]&=&\frac{R^2}{2\pi\alpha'}\int d\tau~\oint d\sigma~\tr\left(\frac12
\sum_{i=1}^8\left(\nabla_{A,\mu}X^i\right)^\dagger\left(\nabla_A^\mu
X^i\right)+\psi^T\nabla\slash_A\psi+g^2F_{A,\mu\nu}F_A^{\mu\nu}\right.\\&
&~~~~~~~~~~~~~~~\left.-\frac1{2g^2}\sum_{i<j}\left[X^i,X^j\right]^2+\frac1g
\sum_{i=1}^8\psi^T\gamma_i\left[X^i,\psi\right]\right)\end{array}}
\label{matrixYMaction}\eeq
where $g$ is the string coupling constant and $\alpha'$ is the string tension.
The bosonic fields $X^i(\tau,\sigma)$, $i=1,\dots,8$, are $N\times N$ Hermitian
matrices in the adjoint representation of the local gauge symmetry group $U(N)$
which transform in the vector ${\bf8}_v$ representation of the global $SO(8)$
R-symmetry group of rotations in the transverse space. They describe the
collective coordinates in ${\bb R}^8$ of the  $N$ D-strings. Their
superpartners are the $N\times N$ fermionic matrices
$\psi(\tau,\sigma)=\pmatrix{\psi_{\rm L}^\alpha(\tau+\sigma)\cr\psi_{\rm
R}^{\dot\alpha}(\tau-\sigma)}$, $\alpha,\dot\alpha=1,\dots,8$, where $\psi_{\rm
L}^\alpha(\tau+\sigma)$ and $\psi_{\rm R}^{\dot\alpha}(\tau-\sigma)$ are
Majorana-Weyl spinor fields in the adjoint representation of $U(N)$ which
transform, respectively, in the spinor ${\bf8}_s$ and conjugated spinor
${\bf8}_c$ representation of $SO(8)$. The trace in (\ref{matrixYMaction}) is
over the $U(N)$ indices. The action $S$ also contains a
residual, non-dynamical two-dimensional $U(N)$ gauge field
$A_\mu(\tau,\sigma)$, $\mu=0,1$, and $F_A=dA+[A,A]/2$ is its curvature. The
world-sheet is the cylinder $(x^0,x^1)\equiv(\tau,\sigma)\in{\bb R}^1\times
S^1$ with $R$ the
radius of the circle $S^1$, so that $\sigma\in[0,2\pi R)$. For a global ${\cal
N}=8$ world-sheet supersymmetry, we need to take the fermion fields to lie in
the Ramond sector of the world-sheet theory, i.e. $\psi^\alpha(\e^{-2\pi
iR}z)=\psi^\alpha(z)$ and $\psi^{\dot\alpha}(\e^{2\pi iR}\bar
z)=\psi^{\dot\alpha}(\bar z)$ where $z=\e^{-i(\tau+\sigma)}$ and $\bar
z=\e^{-i(\tau-\sigma)}$. The gamma-matrices $\gamma^i$ generate the spin(8)
Clifford algebra in a Majorana-Weyl basis. They can be decomposed with respect
to the reducible ${\bf8}_s\oplus{\bf8}_c$ representation of $SO(8)$ as
\beq
\gamma^i=\pmatrix{0&\gamma^i_{\alpha\dot\alpha}\cr\gamma_{\dot\alpha\alpha}^i
&0\cr}
\label{gammai}\eeq
where $\gamma_{\dot\alpha\alpha}^i=(\gamma^i)^T_{\alpha\dot\alpha}$ are
real-valued, and for $i=1,\dots,7$ they are anti-symmetric while
$\gamma^8_{\dot\alpha\alpha}$ is symmetric. The gauge-covariant derivative in
(\ref{matrixYMaction}) is $\nabla_A=d-i[A,~\cdot~]$, and
$\nabla\slash_A=\nabla_A^\mu\Gamma_\mu$ where the world-sheet gamma-matrices
are $\Gamma^0=(\gamma^0)^2=\pmatrix{{\bf1}_8&0\cr0&{\bf1}_8\cr}$ and
$\Gamma^1=\gamma^9=\pmatrix{{\bf1}_8&0\cr0&-{\bf1}_8\cr}$ with ${\bf1}_8$ the
$8\times8$ identity matrix.

The action (\ref{matrixYMaction}) describes a non-trivial matrix field
theory coupled to two-dimensional Yang-Mills theory. The temporal component
$A_0$ of the gauge field is non-dynamical but acts as a Lagrange multiplier
enforcing the invariance of the theory under local gauge transformations.
Varying $S$ with respect to it gives the generator $\cal G$ of these
transformations in the Weyl gauge as
\beq
{\cal G}=4g^2\nabla_A^\sigma(\partial_\tau A_1)-2[X_i,\partial_\tau
X_i]+[\psi^T,\psi]
\label{gaugegen}\eeq
and the constraint ${\cal G}\sim0$ enforces local gauge invariance of the
quantum field theory. It is instructive to consider the model
(\ref{matrixYMaction}) as a bosonic matrix field theory coupled to
two-dimensional adjoint QCD with colour group $U(N)$. There are then eight
flavours of massless adjoint fermions in this interpretation.

Let us start by examining the confinement problem in this model. Consider first
the weak-coupling limit $g^2\to\infty$ of the Yang-Mills theory. In that limit,
the action (\ref{matrixYMaction}) reduces to the ordinary adjoint QCD action
\beq
S_{{\rm QCD}_2}[A,\psi]=\frac{R^2}{2\pi\alpha'}\int d\tau~\oint
d\sigma~\tr\left(\psi^T\nabla\slash_A\psi+g^2F_{A,\mu\nu}F_A^{\mu\nu}\right)
\label{qcd2action}\eeq
plus an additional term for the interaction of the dynamical boson fields
$X^i$ with the Yang-Mills field. In each field sector there are $N^2$ colour
degrees of freedom. As shown in \cite{gkms}, the fundamental path-ordered
Wilson loop
\beq
W_{\cal C}[A]=\frac{\tr_F}{N}~P\exp\left(i\oint_{\cal C}A_\mu(x)~dx^\mu\right)
\label{fundwilsonloop}\eeq
in this case is screened by the massless adjoint fermions, i.e. its vacuum
expectation value in the fermion-gauge sector of the theory obeys a perimeter
law, rather than an area law which is the signal of quark confinement. This
means that a heavy probe charge in the fundamental colour representation, whose
holonomy from parallel transport along the closed contour $\cal C$ due
to its coupling to $A$ is determined by (\ref{fundwilsonloop}), is screened by
the dynamical quarks, in contrast to the confinement phase where one obtains a
linear static potential between external quarks. The classic example of this
phenomenon occurs in the massless
Schwinger model \cite{gkms} which is a $U(1)$ version of (\ref{qcd2action}).
Then the screening of test charges is due to a dynamical Higgs mechanism for
the photon field which induces a non-local Schwinger mass term proportional to
\beq
{\cal S}[A]=\frac1{4g^2}\int d\tau~\oint
d\sigma~\epsilon_{\mu\nu}F_A^{\mu\nu}~\Box^{-1}~\epsilon_{\lambda\rho}
F_A^{\lambda\rho}
\label{schwinger}\eeq
after integrating over the fermion fields. This effect occurs outside a radius
$R_s\sim g^2$. However, in the weak-coupling regime, the screening radius
becomes infinite, and so at $g^2\to\infty$ the matrix string theory is
essentially confining (because the induced dynamical gluon mass vanishes).

The situation is somewhat different in the strong-coupling phase $g^2\to0$.
This regime corresponds to the free string limit and the resulting field theory
(\ref{matrixYMaction}) is quite different from the adjoint QCD which dominates
the weak-coupling phase. At $g^2\to0$ we have $[X^i,X^j]=0$, so that the
D-string coordinates can be simultaneously diagonalized and hence taken to lie
in the Cartan subalgebra of $U(N)$, i.e.
\beq
X^i(\tau,\sigma)={\rm diag}[x^i_1(\tau,\sigma),\dots,x^i_N(\tau,\sigma)]
\label{cartans}\eeq
The D-strings now have well-defined configurations described by the
coordinate fields $x_a^i$ (in contrast to the generic case where the fields
reside in a non-commutative spacetime geometry). There are now only $N$ colour
degrees of freedom in each field sector and the model is described in terms of
the Green-Schwarz light-cone coordinates $x_a^i$, $\psi_a^\alpha$ and
$\psi^{\dot\alpha}_a$ with $a=1,\dots,N$. The effective field theory is the
${\cal N}=8$ supersymmetric sigma-model in the orbifold target space $({\bb
R}^8)^N/S_N$ \cite{dvv}. From the point of view of the Schwinger model, we
effectively have $N$ abelian gauge copies in which all fermion and boson fields
are neutral because they lie in the adjoint representation of the local gauge
group. There are no massless dynamical charged degrees of freedom, and hence no
screening. Furthermore, the gauge symmetry is broken as $U(N)\to U(1)^N$ at
$g^2\to0$ so that an external quark yields a topological sector with the
quantum numbers of $N$ D-strings and one elementary string associated with a
$U(1)$ factor \cite{wittenD}.

Thus at both weak and strong coupling we expect external colour probe charges
to be confined in the quantum field theory (\ref{matrixYMaction}). If we now
apply the analysis of \cite{kss} which suggests that there are no phase
transitions in this theory as the string coupling constant $g$ is continuously
varied, then we can deduce that the matrix field theory (\ref{matrixYMaction})
is confining for all values of $g^2$. This leads to a picture of string states
in the gauged supersymmetric matrix model (\ref{matrixYMaction}) as QCD strings
formed in a confining phase. These matrix strings are hadronic states which
form $SO(8)$ flavour multiplets and $U(N)$ colour singlets. These strings are
not expected to be the usual rigid strings of QCD (except possibly at
weak-coupling), but more likely critical strings.

The hadronic spectrum is actually richer, because we have not yet exploited the
topology of the underlying cylindrical world-sheet which has non-trivial
fundamental group $\pi_1({\bb R}^1\times S^1)={\bb Z}$ that gives the
two-dimensional gauge field $A$ finitely-many topological degrees of freedom.
The locally gauge-invariant Wilson loops (\ref{fundwilsonloop}) are, as in the
confinement discussion above, the crucial objects which identify the
non-perturbative states of the theory. In particular, as discussed in
\cite{kogzhit}, on a cylinder there are now actually two types of hadronic
states in the spectrum, namely those associated with contractible and
non-contractible loops ${\cal C}^{(n)}\in\pi_1({\bb R}^1\times S^1)$ on the
world-sheet in (\ref{fundwilsonloop}) (Fig. 1), where $n\in{\bb Z}$ is the
number of times that the loop winds around the circle $S^1$ of the world-sheet
with a particular orientation. The topologically non-trivial ones ($n\neq0$)
are those associated with the gauge field configurations $A_\mu$ that encircle
the spatial part $S^1$ of the world-sheet cylinder. Their spatial components
admit the Hodge decomposition
\beq
A_1(\tau,\sigma)=\frac1R\sigma+\partial_\tau\chi(\tau,\sigma)+\partial_
\sigma\alpha(\tau,\sigma)
\label{A1hodge}\eeq
In addition to these pure gluon winding modes, there are also the more
complicated locally gauge invariant operators involving the adjoint boson or
fermion fields $\Phi$ (Fig. 1)
\beq\new{\begin{array}{lll}
W_{{\cal C}^{(n)}}^{(x_1,\dots,x_m)}[A,\Phi]
&=&\frac{\tr_F}{N}~P~\Phi(x_1)\exp\left(i\int_{x_1}^{x_2}
A_\mu(x)~dx^\mu\right)\Phi(x_2)\cdots\\& &~~~~~~~\times\cdots\Phi(x_m)
\exp\left(i\int_{x_m}^{x_1}A_\mu(x)~dx^\mu\right)\end{array}}
\label{mixedwilsonloop}\eeq
inserted at points $x_1,\dots,x_m$ on the non-contractible loop ${\cal
C}^{(n)}$.

\bigskip

\begin{figure}[htb]
\centerline{\psfig{figure=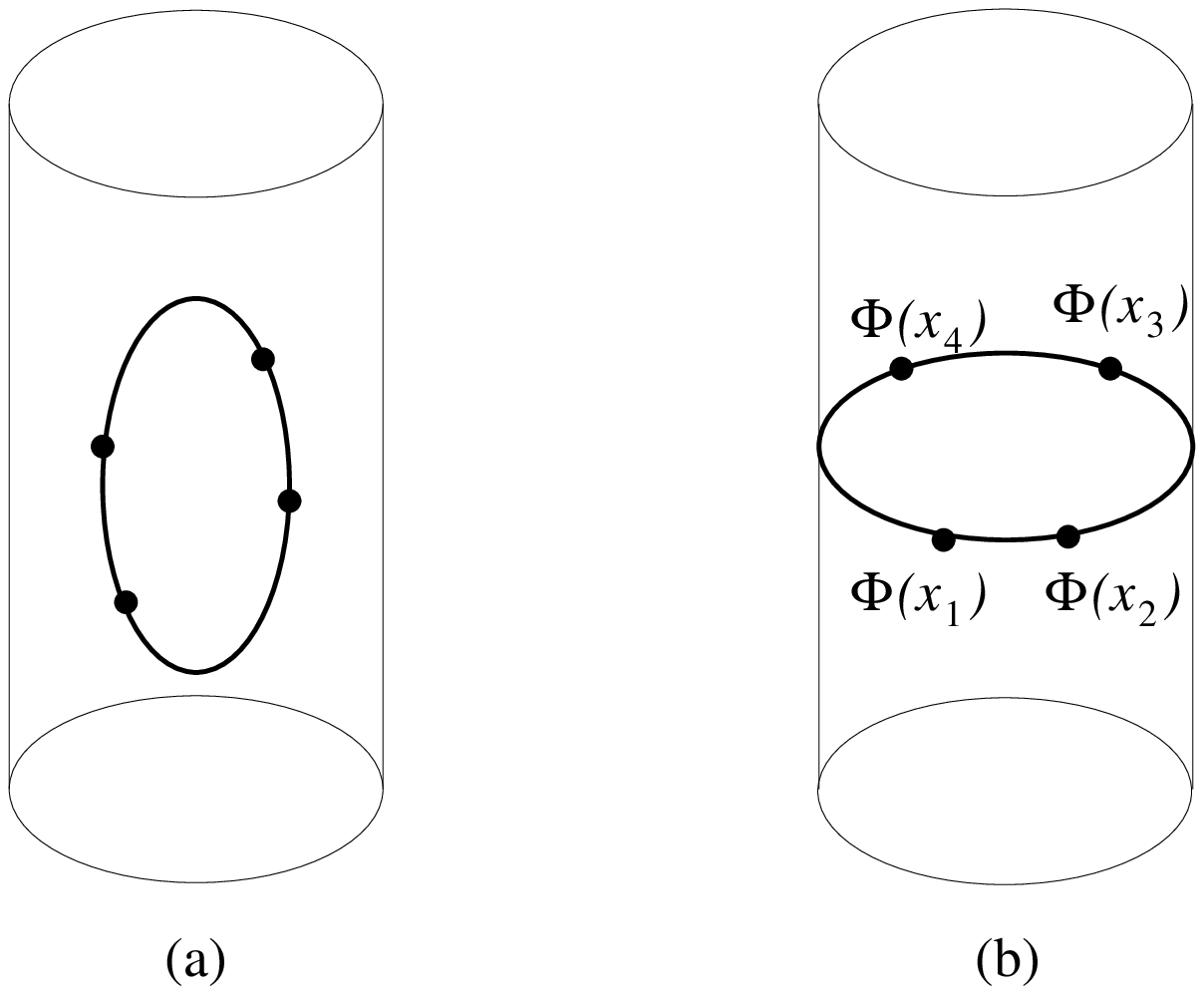,height=0.4\textwidth}}
\begin{description}
\small
\baselineskip=12pt
\item[Figure 1:] The two types of hadronic string states on the matrix
world-sheet ${\bb R}^1\times S^1$. The bold lines denote the gluon strings and
the solid circles depict insertions of the adjoint matter fields along the
Wilson loops. The hadrons propagate between the bottom and the top of the
cylinder and form either contractible loops as shown in (a) or the winding
modes depicted in (b) whereby they encircle the loop of the world-sheet $n$
times. These strings define $N$ inequivalent vacua which are connected by
topological instantons.
\end{description}
\end{figure}

The fundamental Wilson loops (\ref{fundwilsonloop}) are in general not
invariant under large gauge transformations. Under a global gauge
transformation which winds $n$ times around $S^1$, i.e. $\sigma\to\sigma+2\pi
Rn$, the fundamental Wilson loop twists
\beq
W_{{\cal C}^{(n)}}[A]\to\e^{2\pi in/N}~W_{{\cal C}^{(n)}}[A]
\label{wilsontransf}\eeq
by an element in the center ${\bb Z}_N$ of the group $SU(N)$. This property
follows from the fact that the adjoint matter fields are decoupled from the
photon field in the decomposition of the $U(N)$ gauge field $A$ with respect to
the exact sequence
\beq
1\to{\bb Z}_N\to SU(N)\times U(1)\to U(N)\to{\bf1}
\label{UNseqn}\eeq
and it is intimately related to the fact that adjoint QCD has exotic
topological properties. The global gauge group is effectively $SU(N)/{\bb Z}_N$
which has a non-trivial fundamental group
\beq
\pi_1(SU(N)/{\bb Z}_N)=\pi_0({\bb Z}_N)={\bb Z}_N
\label{fundgp}\eeq
and so there are $N$ topologically inequivalent superselection sectors in the
vacuum of the quantum field theory \cite{wittenQCD}. The non-trivial topology
(\ref{fundgp}) of the global gauge group yields $N$ gauge-inequivalent vacuum
states $|0\rangle_n$ labelled by the $N$-th roots of unity \cite{kogzhit}. In
particular, we see the appearence of the usual Yang-Mills instantons, which
interpolate between the inequivalent vacua, as new non-perturbative excitations
in the matrix string theory (\ref{matrixYMaction}). The explicit instanton
field configurations are represented by the Wilson loop operators $W_{{\cal
C}^{(n)}}[A]$ which are interpolating fields for the inequivalent vacua as
described by their twisting property (\ref{wilsontransf}).

Thus, in addition to the local hadronic strings arising from the confining
properties of the theory, there are global inequivalent vacua $|0\rangle_n$
which are described by topological strings which form non-contractible loops
${\cal C}^{(n)}$ of a given winding number $0\leq n<N$. The winding numbers
label the topological class of the parallel transport operators $W_{{\cal
C}^{(n)}}[A]$  which effectively act in the fibers of the principal $SU(N)/{\bb
Z}_N$ bundle over the world-sheet and describe interpolating instanton
configurations between the vacua. In addition to the spectrum of winding modes
corresponding to the pure gluon non-contractible loops, there are also the
mixed Wilson loop operators (\ref{mixedwilsonloop}), with adjoint matter fields
inserted along the non-contractible loops, which can decay into the usual
hadron modes and pure gluon winding modes $W_{{\cal C}^{(n)}}[A]$. Note that
these strings differ from the fundamental strings discussed in
\cite{fundstring} which correspond to the photon field in the supersymmetric
Yang-Mills theory, and also from the twisted long string states in
\cite{motl,dvv}. In the limit $N\to\infty$, the field theory has an infinite
number of degrees of freedom with infinitely many inequivalent vacua
$|0\rangle_n$. Then the vacuum sector of the D-string Fock space contains an
infinity of hadron and gluon string modes with infinitely many instanton
configurations. These objects can be identified as non-perturbative degrees of
freedom in the membrane theory under consideration. It would be interesting to
see if there is any sort of instanton condensation in the ground state upon
taking the large-$N$ limit corresponding to a degeneracy in the twisting
property (\ref{wilsontransf}) of the gluon winding modes.

We have not yet fully exploited the dynamics of the D-string embedding fields
$X^i$ in the eight-dimensional target space (or equivalently the ${\cal N}=8$
supersymmetry of the matrix model (\ref{matrixYMaction})). It turns out, as we
now demonstrate, that there is in fact a much larger class of target space
instanton field configurations which reduce, in the weak-coupling phase, to the
world-sheet instantons of two-dimensional Yang-Mills theory that we described
above. For this, we consider the vacuum amplitude which is given by the path
integral
\beq
Z(R,g;N)=\int DA~DX~D\psi~\e^{-NS[A;X,\psi]}
\label{partfn}\eeq
where it is understood that the functional integration is restricted to gauge
orbits and to fermion field configurations with periodic boundary conditions.
In the case of constant world-sheet field configurations (the ultra-local form
of the theory), the integration over the constant modes $X_0$ and $\psi_0$ of
the matrix degrees of freedom was explicitly carried out in \cite{kss}. There
it was shown that the triality property ${\bf8}_v\cong{\bf8}_s\cong{\bf8}_c$ of
the rotation group $SO(8)$ leads to an explicit non-perturbative Nicolai map
\cite{nicolai,bbrt} associated with the ${\cal N}=8$ supersymmetry. The Nicolai
map
\beq
X_0^i\to
W^i(X_0)=\mbox{$\frac12$}~\gamma^i_{\alpha\dot\alpha}
\left[X_0^\alpha,X_0^{\dot\alpha}
\right]
\label{nicolai0}\eeq
trivializes the bosonic part of the action (\ref{matrixYMaction}) by mapping it
into a Gaussian form for the fields $W^i$. Moreover, the Jacobian determinant
that arises from the field transformation (\ref{nicolai0}) cancels exactly with
the determinant that arises from integrating out the zero modes of the bilinear
fermionic part of (\ref{matrixYMaction}). The integration over the constant
modes in (\ref{partfn}) is thus formally unity. The model
(\ref{matrixYMaction}) defined with only constant field modes included is in
fact the action for the low-energy dynamics of D-instantons \cite{wittenD}
which is obtained by dimensionally reducing 10-dimensional supersymmetric
Yang-Mills theory down to zero dimensions and which has been recently studied
as a non-perturbative model for weakly-coupled type-IIB superstrings \cite{2b}.
By T-duality, the IIA and IIB theories are equivalent, and this equivalence
itself hints at a possible emergence of instanton configurations in the
D-string field theory (\ref{matrixYMaction}).

A standard diagrammatic argument (or the Schwinger-Dyson equations) shows that
the leading contribution to the partition function (\ref{partfn}) at $N=\infty$
can be evaluated as
\beq
\lim_{N\to\infty}Z(R,g;N)=\int DA~\exp\left\{-\frac{NR^2g^2}{2\pi\alpha'}\int
d\tau~\oint d\sigma~\tr~F_{A,\mu\nu}F_A^{\mu\nu}\right\}~\sqrt{{\cal Z}[A]}
\label{partfnA}\eeq
where the path integral
\beq\new{\begin{array}{lll}
{\cal Z}[A]&=&\int D\phi~D\phi^\dagger~D\chi~D\bar\chi~\exp\left\{-\frac{NR^2}
{2\pi\alpha'}\int d\tau~\oint d\sigma~\tr\left[\sum_{i=1}^8\left(\bar
\nabla_A\phi_i^\dagger\right)\left(\nabla_A\phi^i\right)+2\bar\chi
{}~\nabla\slash_A~\chi\right.\right.\\&
&\left.\left.~~~~~-\frac1{g^2}\sum_{i<j}\left[\phi_i^\dagger,\phi^j\right]
\left[\phi^i,\phi_j^\dagger\right]+\frac2g\sum_{i=1}^8\left(\bar\chi^\alpha
\gamma^i_{\alpha\dot\alpha}\left[\phi^i,\chi^{\dot\alpha}\right]+
\bar\chi^{\dot\alpha}\gamma^i_{\dot\alpha\alpha}
\left[\phi_i^\dagger,\chi^\alpha\right]\right)\right]\right\}\end{array}}
\label{calZdef}\eeq
is defined as a functional integration over complex field configurations. We
shall always ignore irrelevant (infinite) numerical factors in what follows.
Here $\phi^i(\tau,\sigma)$ is an $N\times N$ complex matrix field in the
adjoint
representation of the local gauge group $U(N)$ and in the complexified vector
representation ${\bf8}_v\otimes{\bb C}$ of the global R-symmetry group $SO(8)$.
The $N\times N$ fermionic matrix $\chi(\tau,\sigma)$ is a complex, periodic
16-component fermion field in the adjoint representation of $U(N)$ whose chiral
and anti-chiral components transform as Dirac-Weyl spinors in
${\bf8}_s\otimes{\bb C}$ and ${\bf8}_c\otimes{\bb C}$ under $SO(8)$. We have
also introduced the chiral covariant derivatives
$\nabla_A=\partial-i[A_+,~\cdot~]$ and
$\bar\nabla_A=\bar\partial-i[A_-,~\cdot~]$, where
$\partial=\partial_\tau+\partial_\sigma$,
$\bar\partial=\partial_\tau-\partial_\sigma$ and
$A_\pm(\tau,\sigma)=A_0(\tau,\sigma)\pm A_1(\tau,\sigma)$. This doubling of the
number of degrees of freedom in the model (an increase of the world-sheet
supersymmetry from ${\cal N}=8$ to ${\cal N}=16$) leads, as we now show, to an
explicit large-$N$ solution for the amplitude (\ref{partfn}) as an integral
over gauge orbits only.

A field $\lambda$ in the adjoint representation of the gauge group can be
expanded as $\lambda=\lambda_aT^a$, where $T^a$ are the Hermitian generators of
$U(N)$ which are normalized as $\tr~T^aT^b=\frac12\delta^{ab}$ and obey the
commutation relations $[T^a,T^b]=if^{abc}T^c$. Doing so in (\ref{calZdef}), the
integration over the complex fermion fields gives
\beq\new{\begin{array}{lll}
{\cal Z}[A]&=&\int
D\phi~D\phi^\dagger~\det\left\|\Delta_A^{ab}[\phi,\phi^\dagger]\right\|\\&
&\times\exp\left\{-\frac{NR^2}
{2\pi\alpha'}\int d\tau~\oint d\sigma~\tr\left(\sum_{i=1}^8\left(\bar
\nabla_A\phi_i^\dagger\right)\left(\nabla_A
\phi^i\right)-\frac1{g^2}\sum_{i<j}\left
[\phi_i^\dagger,\phi^j\right]\left[\phi^i,\phi_j^\dagger\right]\right)
\right\}\end{array}}
\label{fermintout}\eeq
where $\Delta_A^{ab}[\phi,\phi^\dagger]$ is the matrix differential operator
\beq
\Delta_A^{ab}[\phi,\phi^\dagger]=\pmatrix{\nabla
_A^{ab}\delta_{\alpha\beta}&\mbox{$-\frac
igf^{abc}\sum_i\gamma^i_{\alpha\dot\alpha}\phi_{i,c}^*$}
\cr\mbox{$-\frac igf^{abc}\sum_i\gamma^i_{\dot\alpha\alpha}\phi^i_c$}&\bar
\nabla_A^{ab}\delta_{\dot\alpha\dot\beta}\cr}
\label{Fabdef}\eeq
and the ultra-local functional determinant in (\ref{fermintout}) is over both
the $U(N)$ indices and the chiral-antichiral indices of spin(8). For this
extended model with ${\cal N}=16$ supersymmetry, the Nicolai map is defined by
the non-analytic field transformation
\beq
\xi^i=\nabla_A\phi^i+\mbox{$\frac1{2g}$}~\gamma^i_{\alpha\dot\alpha}
\left[\phi^{\dagger\alpha},\phi^{\dagger\dot\alpha}\right]~~~~~~,~~~~~~
\xi_i^\dagger=\bar \nabla_A\phi_i^\dagger-\mbox{$\frac1{2g}$}~
\gamma^i_{\dot\alpha\alpha}\left[\phi^{\dot\alpha},
\phi^\alpha\right]
\label{nicolaixi}\eeq
Then the Jacobian of this transformation is $\frac{\delta(\xi_a,\xi_a^*)}
{\delta(\phi_b,\phi_b^*)}=\Delta_A^{ab}$ and the path integral
(\ref{fermintout}) becomes
\beq
{\cal Z}[A]=\int D\xi~D\xi^\dagger~\det\|\Delta_A^{ab}\|\cdot|\det\|\Delta_A
^{ab}\||^{-1}\exp\left\{-\frac{NR^2}{2\pi\alpha'}\int d\tau~\oint
d\sigma~\tr~\sum_{i=1}^8\xi_i^\dagger\xi^i\right\}
\label{ZAtrivial}\eeq
The trivialization of the bosonic action in (\ref{fermintout}) follows from the
appropriate symmetry properties of the spin(8) gamma-matrices in (\ref{gammai})
(see \cite{kss} for details).

However, the Gaussian integration in (\ref{ZAtrivial}) is not completely
trivial, because it does depend on the winding number (or degree) of the
Nicolai mapping (\ref{nicolaixi}), i.e. on the number of distinct
configurations of the original fields which are mapped into a given
configuration of the free Gaussian fields with their algebraic multiplicity.
Since $\xi^i=\xi_i^\dagger=0$ for the trivial field configuration
$\phi^i=\phi_i^\dagger=0$, the number of times that one covers $\xi$-space as
$\phi$ runs through its range can be calculated by following the zeroes of
$\xi^i,\xi_i^\dagger$, which define the instanton and anti-instanton equations
\beq
\nabla_A\phi_{(I)}^i+\frac1g\frac{\delta}{\delta\phi_{(I)}^i}F[\phi_{(I)},
\phi_{(I)}^\dagger]=0~~~~~~,~~~~~~\bar
\nabla_A\phi_{(I),i}^\dagger-\frac1g\frac{\delta}
{\delta\phi^\dagger_{(I),i}}F^\dagger[\phi_{(I)}^\dagger,\phi_{(I)}]=0
\label{instantoneq}\eeq
where the pre-potential is \cite{kss}
\beq
F[\phi,\phi^\dagger]=\int d\tau~\oint
d\sigma~\sum_{i=1}^8\gamma^i_{\alpha\dot\alpha}~\tr~\phi^i
\left[\phi^{\dagger\alpha},\phi^{\dagger\dot\alpha}\right]
\label{prepot}\eeq
Note that the instanton configurations also solve the classical equations of
motion
\beq
\bar\nabla_A\nabla_A\phi_{\rm
cl}^i+\frac1{g^2}\sum_{i,j=1}^8\left(\gamma^i\gamma^j\right)_{\dot\alpha\dot
\beta}\left[\phi_{\rm cl}^{\dagger\dot\alpha},\phi_{\rm
cl}^{\dot\beta}\right]\phi_{\rm cl}^j=0
\label{classeqns}\eeq
which follow from varying the bosonic action in (\ref{calZdef}) with respect to
$\phi^i$ (and similarly for the anti-instanton configurations).

The path integral (\ref{ZAtrivial}) calculated about $\xi^i=\xi_i^\dagger=0$
therefore yields an oriented sum over the different classical
instanton-antiinstanton configurations,
\beq\new{\begin{array}{lll}
{\cal Z}[A]&=&\sum\!\!\!\!\!\!\!\!\int_{~~\phi_{(I)},\phi_{(I)}^\dagger}~{\rm
sgn}\det\left\|\Delta_A[\phi_{(I)},\phi_{(I)}^\dagger]\right\|\\&=&
\sum\!\!\!\!\!\!\!\!\int_{~~\phi_{(I)},\phi_{(I)}^\dagger}~\exp\left
\{\mbox{$\frac{\pi i}2$}\left(\zeta(\Delta_A[
\phi_{(I)},\phi_{(I)}^\dagger])-\eta(\Delta_A[\phi_{(I)},\phi_{(I)}^\dagger])
\right)\right\}\end{array}}
\label{calZsum}\eeq
where
\beq\new{\begin{array}{l}
\zeta(\Delta_A)=\lim_{s\to0}~\frac1{\Gamma(s)}\int_0^\infty
dt~t^{s-1}~\tr\|\e^{-t|\Delta_A|}\|
\\\eta(\Delta_A)=\lim_{s\to0}~\frac1{\Gamma(\frac{s+1}2)}\int_0^
\infty dt~t^{(s-1)/2}~\tr\|\Delta_A~\e^{-t\Delta_A^2}\|\end{array}}
\label{zetaetadef}\eeq
are, respectively, the Riemann zeta-function and Atiyah-Patodi-Singer
eta-invariant which measure the spectral volume and spectral asymmetry of the
self-adjoint first-order differential operator $\Delta_A$. The functional
(\ref{calZsum}) is either a sum over discrete instanton-antiinstanton
configurations or an integral over the instanton-antiinstanton moduli space
${\cal M}_I$ (the solutions to (\ref{instantoneq}) modulo local gauge
transformations and global R-symmetry transformations), and it is understood
that zero-modes of $\Delta_A$ are excluded. Thus the path integral
(\ref{calZdef}) represents a topological invariant of the cylindrical
worldsheet embedding in ${\bb R}^8$, and, in particular, it is semi-classically
exact. This property is a result of the usual non-renormalizations in
supersymmetric field theories, and it is also the generic situation in a
topological field theory \cite{bbrt}. The $N=\infty$ vacuum amplitude of the
matrix string theory can thus be written as that of an effective
two-dimensional gauge theory with a local Yang-Mills kinetic term and an
additional non-local term for the gauge fields,
\beq\new{\begin{array}{l}
\lim_{N\to\infty}Z(R,g;N)\\~~~~~~=\sum\!\!\!\!\!\!\!\!\int_{~~{\cal M}_I}~\int
DA~\exp\left\{-\frac{NR^2g^2}{2\pi\alpha'}\int d\tau~\oint
d\sigma~\tr~F_{A,\mu\nu}F_A^{\mu\nu}+\frac{\pi
i}4\left[\zeta(\Delta_A)-\eta(\Delta_A)\right]\right\}\end{array}}
\label{gaugetheff}\eeq

We note that in the weakly-coupled regime of the underlying two-dimensional
Yang-Mills theory in (\ref{partfnA}) ($g^2\to\infty$), the instanton equations
(\ref{instantoneq}) become the usual ones $\nabla_A\phi_{(I)}=\bar
\nabla_A\phi_{(I)}^\dagger=0$, which are solved by a path-ordered Wilson line
(so that $\phi\equiv\epsilon_{\mu\nu}F_A^{\mu\nu}$) and yield the classical
field equations $\bar\nabla_A\nabla_A\phi_{\rm
cl}=\nabla_A\bar\nabla_A\phi_{{\rm cl}}^\dagger=0$. The Hessian which appears
in (\ref{calZsum}) is then a positive operator and so the path integral ${\cal
Z}[A]$ just counts the total number $N_I=\dim{\cal M}_I$ of ordinary Yang-Mills
instantons of the given principal $U(N)$-bundle over the world-sheet cylinder
${\bb R}^1\times S^1$. Thus the effect of the path integral (\ref{calZdef}) is
to add to the Yang-Mills action in (\ref{partfnA}) a topological instanton term
and we recover the world-sheet instanton configurations along with a sum over
$\theta$-vacua (with $\theta\in{\bb Z}_N$) that we described earlier. It is
intriguing that, since the ordinary Yang-Mills action in two-dimensions is a
topological BF field theory in the limit of weak coupling \cite{bbrt}, the full
supersymmetric matrix model defines a topological field theory at
$g^2\to\infty$. Thus in the weak-coupling phase the we encounter a purely
topological field theory with only global degrees of freedom.

In the strong-coupling limit of the two-dimensional Yang-Mills theory
($g^2\to0$), the instanton configurations are defined by the extrema of the
pre-potential (\ref{prepot}). Then the two-dimensional matrix model is
effectively independent of the gauge field $A$ and it coincides essentially
with the reduced model studied in \cite{kss}. This reduced model was shown to
reproduce many of the characteristic features of the matrix quantum mechanics
describing the short-distance properties of D0-branes in weakly-coupled
type-IIA superstring theory \cite{bfss}. Again the effective theory thus
obtained is essentially a topological field theory, consistent with the fact
that in the free string limit the matrix model reduces to a two-dimensional
supersymmetric sigma-model \cite{taylor,dvv,bbrt}.

For finite $g^2$, the instanton equations (\ref{instantoneq}) yield a
one-parameter family of instanton field configurations which interpolate
between ordinary Yang-Mills instantons at $g^2\to\infty$ (connecting the
different vacua of the effective two-dimensional adjoint QCD) and the
configurations which minimize the pre-potential (\ref{prepot}) describing the
characteristics of the matrix model for M theory in the decompactified limit
$R^2\to\infty$ \cite{bfss}. These interpolating instantons are non-perturbative
configurations in the gluon sector of the theory which can form new sorts of
glueball states. These states are new types of non-perturbative degrees of
freedom in the full 11-dimensional M theory. The relevence of instanton quantum
numbers has been noted in \cite{fhrs} for the case where the M theory matrix
model is compactified on a 4-torus $(S^1)^4$. There the instanton charge is
identified with the wave-number of the photon field in the supersymmetric
Yang-Mills theory along a new fifth direction in the quantum field theory. For
a general compactification on an $n$-torus $(S^1)^n$, the type-IIA strings are
$(n-1)$-dimensional domain walls wrapping around a cycle of the T-dual torus
\cite{fundstring}. In our case, in the large-$N$ limit we obtain infinitely
many such solitonic walls from the $\theta$-vacua that we described before, and
the corresponding world-sheet instantons are then mapped into the target space
${\bb R}^8$ by the instanton fields $\phi_{(I)}^i$. It is unclear though what
the happens to the structure of these ${\bb Z}_N$ domains at $N\to\infty$. The
instantons represent non-perturbative membrane excitations in the transverse
space, and it would be interesting to see if they correspond to any previously
observed M theoretic membranes. In particular, it would be interesting to see
if these $\theta$-vacua have anything to do with the construction of
longitudinal membranes of the matrix model \cite{berdoug,taylor}. Indeed, the
physics of wrapped membranes (and other non-perturbative degrees of freedom)
may be completely contained in the sum (\ref{gaugetheff}) over the different
topological vacuum sectors of the matrix supersymmetric Yang-Mills theory. We
note again that the strings that these configurations represent are not the
usual ones obtained from the abelian sector of the theory, but are associated,
in a non-trivial way via the matrix matter fields, with the topology of the
full principal fiber bundle of the gauge theory.

Let us stress that the final result (\ref{gaugetheff}) for the vacuum amplitude
is not as simple as it seems at first glance. First of all, the Nicolai map
that we have exhibited is only valid at the leading order of the $1/N$
expansion of the partition function. Thus the instanton sum (\ref{gaugetheff})
represents a localization of the vacuum amplitude onto the instanton moduli
space associated with the large-$N$ group theory of $U(\infty)$ or
$SU(\infty)$. For the full theory at finite $N$, the model is highly
non-trivial. Indeed, Nicolai maps in supersymmetric gauge theories are only
known in general at a perturbative level because of the highly non-local
character of the mapping \cite{nicolai}. The only non-trivial cases where they
can be constructed exactly are some low-dimensional models and topological
field theories \cite{bbrt}, in which case they localize the path integrals onto
finite-dimensional objects which are topological invariants of the underlying
spacetime. The large-$N$ limit that we have taken above effectively achieves
this property. The second problem is that at this stage we have no idea as to
the general structure of the instanton moduli space ${\cal M}_I$, or the
related problem of inverting the Nicolai map to compute non-supersymmetric
correlators of the matrix model at large-$N$. It is not immediate how the
classification of these target space instantons will carry through in terms of
the topology of the principal $U(N)$-bundle (for the world-sheet Yang-Mills
instantons, they are classified by the usual first Chern cohomology classes
$c_1(U(N))=F_A/2\pi$). The entire non-triviality of the theory rests in these
two important aspects of the large-$N$ theory \cite{kss}, and it would be
interesting to deduce more precisely what the topological instantons above
represent. Another important problem is using the Nicolai mapping to determine
if there is a non-zero vacuum chiral condensate which may then classify the
different vacuum states by ${}_n\langle0|\psi^T\psi|0\rangle_n\sim\e^{2\pi
in/N}$ \cite{kogzhit}, as they do in four-dimensional supersymmetric Yang-Mills
theory. It would also be interesting to determine the properties of Wilson loop
observables in this effective gauge theory, as well as other non-supersymmetric
observables constructed from the dimensionally-reduced fields themselves thus
making contact with objects such as D-particles and D-membranes \cite{dvv}.

As we have discussed, ordinary Yang-Mills theory in two-dimensions is
essentially a topological field theory with Nicolai map $\xi(A)=F_A^{01}$
\cite{bbrt}, and its partition function receives contributions only from the
moduli space of flat connections on the world-sheet. This reflects the
relationship between topological Yang-Mills theories and conformal field theory
\cite{bbrt}. The result (\ref{gaugetheff}) is an extension of this fact to the
matrix string theory, and it would be interesting to see how this effective
gauge theory localizes onto a finite-dimensional object, given that it still
represents a theory with only global degrees of freedom. It would also be
interesting to connect the theory (\ref{gaugetheff}) with the Gross-Taylor
expansion of two-dimensional Yang-Mills theory \cite{grosstaylor}. This was
achieved starting from the heat-kernel expansion of the partition function for
two-dimensional Yang-Mills theory \cite{rusakov}
\beq
Z_{\Sigma_T}(h,A)=\sum_R~(\dim R)^{2-2h}~\e^{-A(\Sigma_T)C_2(R)/2N}
\label{YM2}\eeq
in a target space $\Sigma_T$ which is a compact Riemann surface of genus $h$
and total area $A(\Sigma_T)$. Here the sum is over unitary irreducible
representations $R$ of the gauge group $U(N)$ and $C_2(R)$ is the second
Casimir of $R$. The Gross-Taylor series expands the exact solution (\ref{YM2})
in $1/N$ and interprets the individual terms as string maps into $\Sigma_T$. In
this case, in order to maintain the string picture at genus $h\neq1$ one needs
to also introduce extra points on the surfaces which are weights associated
with the Euler character of the Hurwitz moduli space. It would be interesting
to see if the non-local additions to the Yang-Mills action in
(\ref{gaugetheff}) corresponds in any way to the weightings by these
topological invariants, thereby providing a relatively simple matrix model
representation of the sum over covering maps of the Hurwitz moduli space. It
would also be interesting to determine if this topological term removes the
rigidity of the usual QCD strings obtained in this way, leading potentially to
a statistical sum over {\it critical} string mappings. The verification of
these features requires a better understanding of the instanton moduli space
${\cal M}_I$ obtained above in order to find the leading orders in $1/N$ from
(\ref{gaugetheff}).

\end{document}